# How do electronic carriers cross Si-bound alkyl monolayers?

Adi Salomon[#], Olga Girshevitz[#], David Cahen[#*], Till Bocking[≠], Calvin K. Chan[+], Fabrice Amy[+], and Antoine Kahn[+]

[#]Department of Materials and Interfaces, Weizmann Institute of Science, Rehovot, Israel,

[≠]Department of Physics, University of New South Wales, Sydney, NSW, Australia and

[+]Department of Electrical Engineering, Princeton University, Princeton NJ, USA

Electron transport through Si-C bound alkyl chains, sandwiched between n-Si and Hg, is characterized by two distinct types of barriers, each dominating in a different voltage range. At low voltage, current depends strongly on temperature but not on molecular length, suggesting transport by thermionic emission over a barrier in the Si. At higher voltage, the current decreases exponentially with molecular length, suggesting tunneling through the molecules. The tunnel barrier is estimated, from transport and photoemission data, to be ~1.5 eV with a $0.25m_e$ effective mass.

A key issue in molecular electronics is the nature of electron transport through the organic molecules. Are multiple transport mechanisms or a single one operating and are these mechanisms fundamentally new or known already in non-molecular systems? Experimental results are often analyzed in terms of multiple mechanisms, and it is crucial to establish if this is justified or if, in doing so, new mechanisms are ignored. The situation is complicated by the strong dependence of the current through the molecular junction on the nature of the chemical bond to the electrodes.[1-4] Molecules, chemically bound to the semiconductor, are likely to affect the energy and density of surface states and, therefore, the semiconductor band bending.[5] Furthermore, the presence of a dipolar layer at the semiconductor/metal interface can affect the barrier for electron transport inside the semiconductor.[6, 7]

We show here that two mechanisms for electron transport occur in *n-Si/alkyl monolayer/Hg* junctions, where the molecules are bound via covalent Si-C bonds to Si. The two mechanisms are unambiguously distinguished and identified, as each dominates in a different range of applied voltage. We chose this semiconductor/molecule/metal system to optimize the possibility of observing current transport over the semiconductor/metal barrier and tunneling through the insulating organic monolayer barrier. There are several key features of the experimental system that allow one to carry out the desired experiments of current transport through molecules. First, the absence of an oxide between Si and the alkyl chain allows the direct study of the electrical properties of the monolayer itself. Second, Si presents a very smooth, reproducible, and well-controlled solid surface.[8, 9] Third, well-prepared Si-C bonded alkyl monolayers are very stable.[10] and suitably long alkyl chains form well-organized, nearly pinhole-free monolayers.[10, 11, 12] .

Fourth, the use of a metallic liquid Hg contact decreases the probability of formation of shorts in any remaining pinhole because of the high surface tension of liquid Hg.[13-15] Finally, the highly covalent Si-C bond, unlike the metal-thiolate bond, is a natural continuation of the alkyl chain. This feature minimizes potential discontinuities at the contacts.[2, 16]. The choice of Si as electrode also makes the system relevant to Si-based microelectronics.

Based on earlier data, we know that binding alkyl molecules to n-Si makes the ohmic, molecule-free, n-Si:H/Hg contact rectifying.[17, 18] Assuming that (i) rectification is due to a Schottky-type metal/semiconductor barrier; and (ii) tunneling across the alkyl monolayer is so efficient that it does not limit the current at low applied voltage, where the Schottky barrier inside the semiconductor is still large, tunneling becomes the limiting step as the barrier decreases with increasing forward bias.[19] We test these assumptions by measuring and analyzing J-V curves as function of temperature, using long chain alkyl molecules with different chain lengths in *n-Si/alkyl monolayer/Hg* junctions.

Si-C linked alkyl monolayers were formed by thermally induced hydrosilylation of alkenes with Si:H, adapting a procedure described elsewhere.[11, 20] Si(111) substrates (1-10 Ω.cm) were rinsed with organic solvents, blown dry under argon, cleaned in piranha solution (conc. $H_2SO_4$ : 30% $H_2O_2$, 3:1 v/v) at 90ºC for at least 20 min and rinsed with DI water (this procedure is essential to get high-quality monolayers, needed for reproducible electrical measurements). Samples were H-terminated by etching in 40% deoxygenated $NH_4F$ solution for 15–20 min. The resulting Si:H surfaces were immersed in the deoxygenated neat alkene under inert atmosphere and heated to 200 °C for at least 4 hr. Control samples of monolayers on Si were characterized by AFM, advancing water contact angle (CA), multi-wavelength, variable angle ellipsometry, Kelvin probe, X-ray Photoelectron Spectroscopy (XPS) and X-ray reflectivity. Empirically, we find the CA to be the simplest, fastest and most reliable way to assess monolayer quality, i.e., highest (lowest) current densities are observed for junctions with lowest(highest) CA values.[21] Only monolayers with CA >105° were used for electrical measurements. Si-C linked molecular monolayers on Si were further analyzed by UV and inverse photoemission spectroscopy (UPS, IPES) to determine the position of the highest occupied and lowest unoccupied molecular orbitals (HOMO, LUMO) after monolayer formation. These measurements were followed by XPS measurements to assess any electron beam damage. Separate XPS measurements served to estimate band bending. The minor traces of O, found by analyzing the C1s, O1s and Si 2p core levels, were identified, based on the energies of these levels, as adventitious $O_2$ adsorbed on the monolayer's outside, i.e., oxide build-up at the Si-monolayer interface is insignificant, even on samples exposed to air for weeks.

In Fig. 1 we show the log(*J*)–*V* characteristics of the series of n-Si/alkyl monolayer/Hg junctions.[22] Their shape clearly differs from earlier results on current transport through Si/alkyl chains /Hg junctions, which were described by thermionic emission with tunneling correction, without distinction between the mechanisms.[18, 23-25] While at low bias the log(*J*)-*V* curves for our n-Si-$C_nH_{2n+1}$//Hg (n=12, 14, 16, 18) junctions are well-described by transport over a Schottky barrier, at higher forward bias (>0.6 V) there is a clear transition to a regime best described by tunneling,[26] i.e., there are two distinct types of barriers for current transport in the system. At low forward biases where current is limited by transport over the barrier inside the Si, the data clearly show that neither monolayer thickness nor molecular length play a discernible role in charge transport. However, at higher forward bias, the observed current depends on molecular length, as expected for tunneling.[27, 28] Such a transition agrees with simulations by Shewchun *et al.* for a similar system, with a transparent conductor instead of Hg and $SiO_x$ as insulator with varying thickness.[29, 30] Their simulations also show that the insulating layer thickness dictates in how far the J-V characteristics deviate from those expected for transport over a Schottky barrier. Indeed, we find that the thicker the monolayer, the lower the voltage at which the current becomes tunneling-limited (cf. Fig. 1). Remarkably, the theoretically predicted behavior was difficult to observe with insulating Si oxide films,[31, 32] as we could confirm for Hg/$SiO_x$/n-Si junctions.[33] The reason is that the density of mobile charges in the oxide also scales with oxide thickness, a complication that is not encountered in the oxide-free system investigated here, in which the insulator is a molecular monolayer.[34, 35]

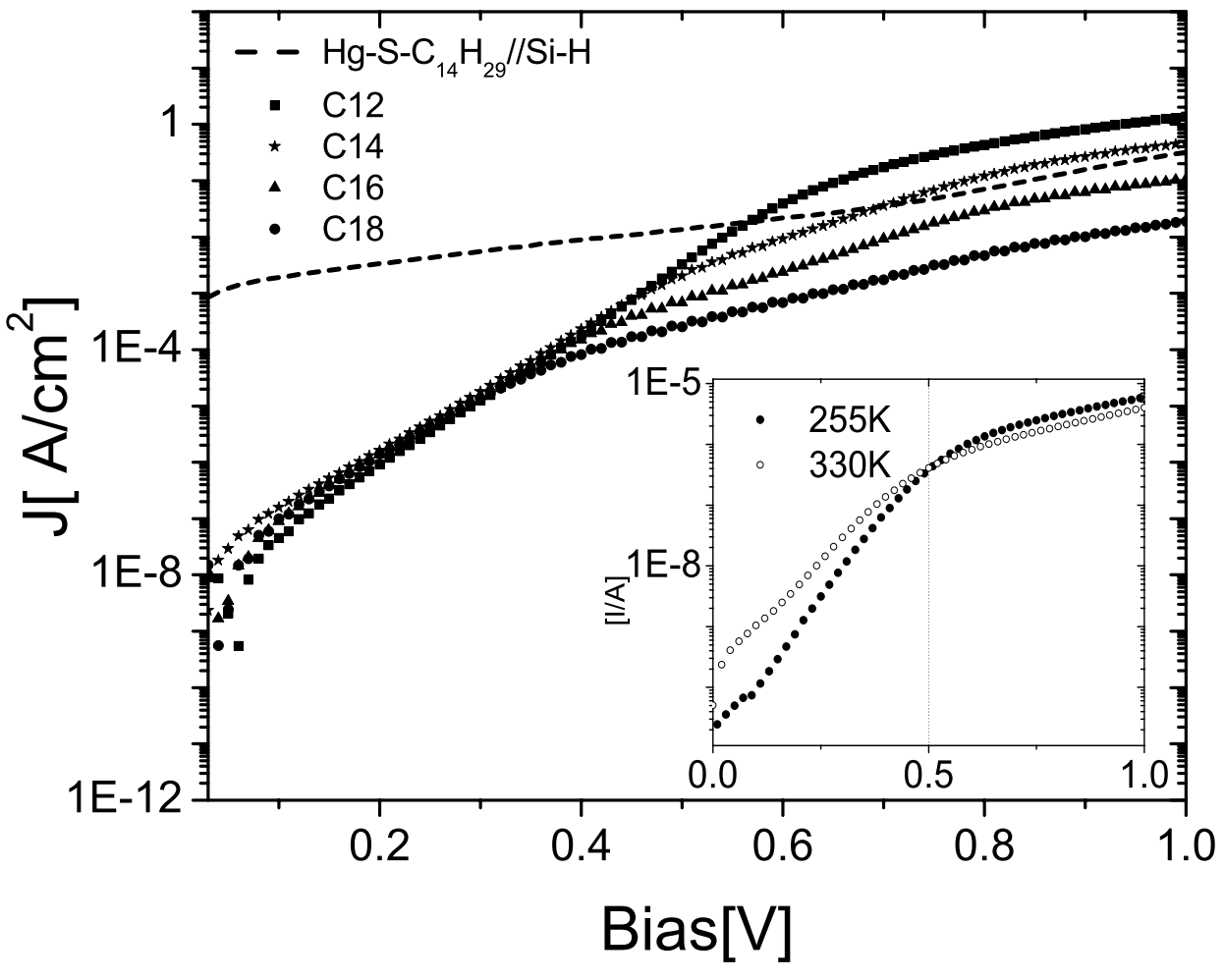

**Figure 1: Experimental J-V curves for n-Si-$C_nH_{2n+1}$//Hg (n=12, 14, 16, 18) junctions at forward bias. Dashed line: n-Si:H//$H_{29}C_{14}$-S-Hg junction. Contact area: 0.1 $mm^2$. Curves are the average for ~ 20 junctions. The error is 10%. Inset: Temperature dependence of I-V curves for n-Si-$C_{14}H_{29}$//Hg over the same voltage span.[36, 37]**

The interpretation of the room temperature J-V characteristics was checked by temperature-dependent I-V measurements. Since Hg solidifies at ~235K, measurements were done from 250K up, with a high temperature limit of 330K because of experimental limitations and toxicity of Hg vapors. We find that, at low voltage, the current increases exponentially with temperature, as expected for current transport over a Schottky barrier.[36] At high voltages the current even decreases slightly with increasing temperature (inset Fig. 1). We attribute this to a change in molecular tilt angle with temperature, as predicted by computations.[37, 38]

The behavior of the n-Si-$C_nH_{2n+1}$//Hg systems can be compared to that of the similar oxide-free n-Si:H//$H_{2n+1}C_n$-S-Hg one (Fig. 1, dashed line), where the chains are chemically bound to Hg via the Hg-S bond and detached from the Si electrode, rather than chemically bound to the Si and detached from the Hg, as n-Si-$C_{14}H_{29}$//Hg.[23] As the n-Si:H//$H_{29}C_{14}$-S-Hg system has a negligible Schottky barrier inside the semiconductor,[39] current through that junction is limited by tunneling only. Indeed, at low voltages the current through the n-Si:H//$H_{29}C_{14}$-S-Hg junction is several orders of magnitudes higher than that through the n-Si-$C_{14}H_{29}$//Hg junction, where the current is limited by thermionic emission over the barrier inside the semiconductor. However, at higher voltages where charge transport is limited in both junctions by tunneling, the two J-V curves converge.

From the experimental data we can extract separately the barrier inside the semiconductor, $\phi_b$, and the barrier for tunneling through the molecular layer, $\phi_t$. $\phi_b$ is calculated using the thermionic emission, or the more general thermionic emission-diffusion model.[36] In all cases, at room temperature and for applied bias, V> ~75 mV the following holds:

$$J \propto \exp(-\frac{q(n\phi_b - V)}{nkT}) \qquad (1)$$

where J is the measured current density, q is the electron charge, and k is Boltzmann's constant.[23] From the slope of the ln(J) - V plot, we find the ideality factor n ~1.5 and from the intercept of this plot, $\phi_b$ ~ 0.8 V. As the effective barrier, ($n\phi_b$ - $V$), decreases with increasing V, the effect of $\phi_b$ decreases and tunneling through the organic molecular layer becomes the limiting factor.

To describe the current in the tunneling-limited regime, we use Simmons' model for tunneling through a rectangular barrier, where the observed current depends exponentially on $d$, the tunneling distance, and on $\beta$, the inverse tunneling decay length:[27, 28]

$$J = J_o e^{-\beta d} \qquad (2)$$

For a rectangular barrier, $\beta$ is expected to depend on applied voltage according to:[27, 28]

$$\beta = 4\pi\sqrt{\frac{2m^*(\phi_t - \frac{qV}{2})}{h^2}} \qquad (3)$$

where $m^*$ is the electron effective mass and $h$ is the Plank constant. From equation (3), $\beta$ should decrease as $V$ increases. While experimentally no dependence[40] or an opposite one is often found,[23] the system studied here shows the expected dependence, as illustrated by the linear $\beta^2$ - $V$ plot in Fig.2.

Such dependence was also reported for alkyl monolayers attached on one side to nm-sized Au contacts.[41] By extrapolating the $\beta^2$ - $V$ plot to V=0, we can evaluate $\phi_t$ and $m^*$ for tunneling through the molecular insulator. The slope of the $\beta^2$ - $V$ plot yields $m^*$ =0.25±0.2 $m_e$, similar to some experimental results and theoretical calculations.[41-43] With this value and the $\beta^2$ - $V$ plot intercept, we find $\phi_t$ = 1.5±0.5 eV. We stress that these $\phi_t$ and m* values are not absolute ones because of uncertainties regarding barrier shape and junction symmetry, factors that are included in m* in the theoretical model.[41] [2]

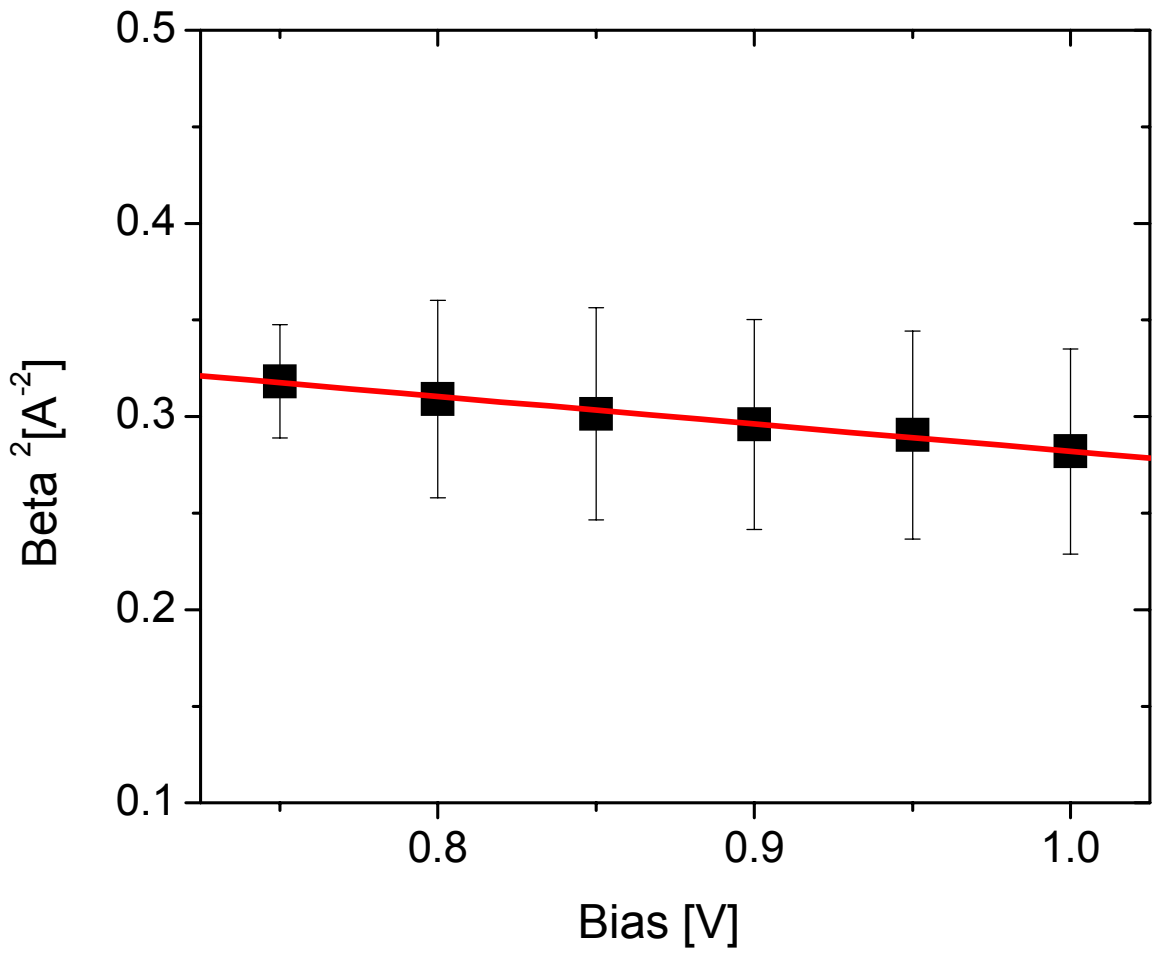

:

**Figure 2: $\beta^2$ vs. forward bias at high voltages, derived from the data shown in Fig. 1 and equations (2) and (3). The solid line shows the best linear fit.**

To extract absolute values for $\phi_t$ and m* we need an independent measurement of one of these quantities. The barrier height for electron transport through the LUMO was obtained from IPES measurements[44] on the n-Si-C14 and n-Si-C16 systems. These show the LUMO at 2.2±0.3 eV above the Fermi level ($E_F$), yielding a tunneling barrier between the conduction band edge ($E_{CB}$) and the LUMO[36] of ~1.4±0.3 eV, after taking into account that the Schottky barrier, $E_{CB}$ - $E_F$ (=$\phi_b$) is ~0.8 eV. The 1.4 eV value is in good agreement with $\phi_t$ derived from the transport data. We note that the low m* (=0.25$m_e$) implies that the *effective* barrier for tunneling is much lower than $\phi_t$.

The band diagram for the n-Si-$C_nH_{2n+1}$//Hg system is constructed with the LUMO position given above, the Schottky barrier found from the I-V measurements, and a 1.1 eV Si band gap (Fig. 3). The HOMO position, determined by UPS, is 3.2 ±0.3 eV below the Fermi level, or 2.9±0.3 eV below the valence band edge ($E_{CB}$ - $E_F$ ≅ 0.8 eV). The HOMO-LUMO gap is then estimated at 5.4 ± 0.5 eV, significantly less than the 9 - 10 eV free alkyl chain value.[42] The reduction in ionization energy and increase in electron affinity for the condensed molecule is, to a great extent, due to stabilizing effect of electronic polarization in the metal or dielectric substrate and in the molecular layer[45]. Furthermore, in a molecular orbital picture, chemical bond formation between the Si surface and the free molecules involves new orbitals for the (*Si+molecule*) system, resulting from interactions between the molecule frontier orbitals and the semiconductor surface state levels (cf. ref.[5]), leading to a HOMO-LUMO gap of the (*Si+molecule*) system that is smaller than that of the free molecule. The HOMO level will be mainly localized on the Si-C bond part of the system.

In summary, our results show that two distinct electron transport mechanisms operate in the n-Si-$C_nH_{2n+1}$//Hg molecular system. At low forward bias voltages, current is limited by transport over a Schottky barrier in the semiconductor. At higher forward bias current is limited by tunneling through the organic molecular layer. The thinner the molecular layer, the higher the voltage at which the tunnel current starts to dominate.[46, 47]

The molecules in the n-Si-$C_nH_{2n+1}$//Hg junction are more ideally insulating than native Si oxide. This may be of interest, for example, in solar cell applications.[30] By using more sophisticated molecules and combining the alkyl chain framework with di-polar molecules,[6] it should be possible to achieve molecular control over both transport via tunneling and transport over the barrier in the system.

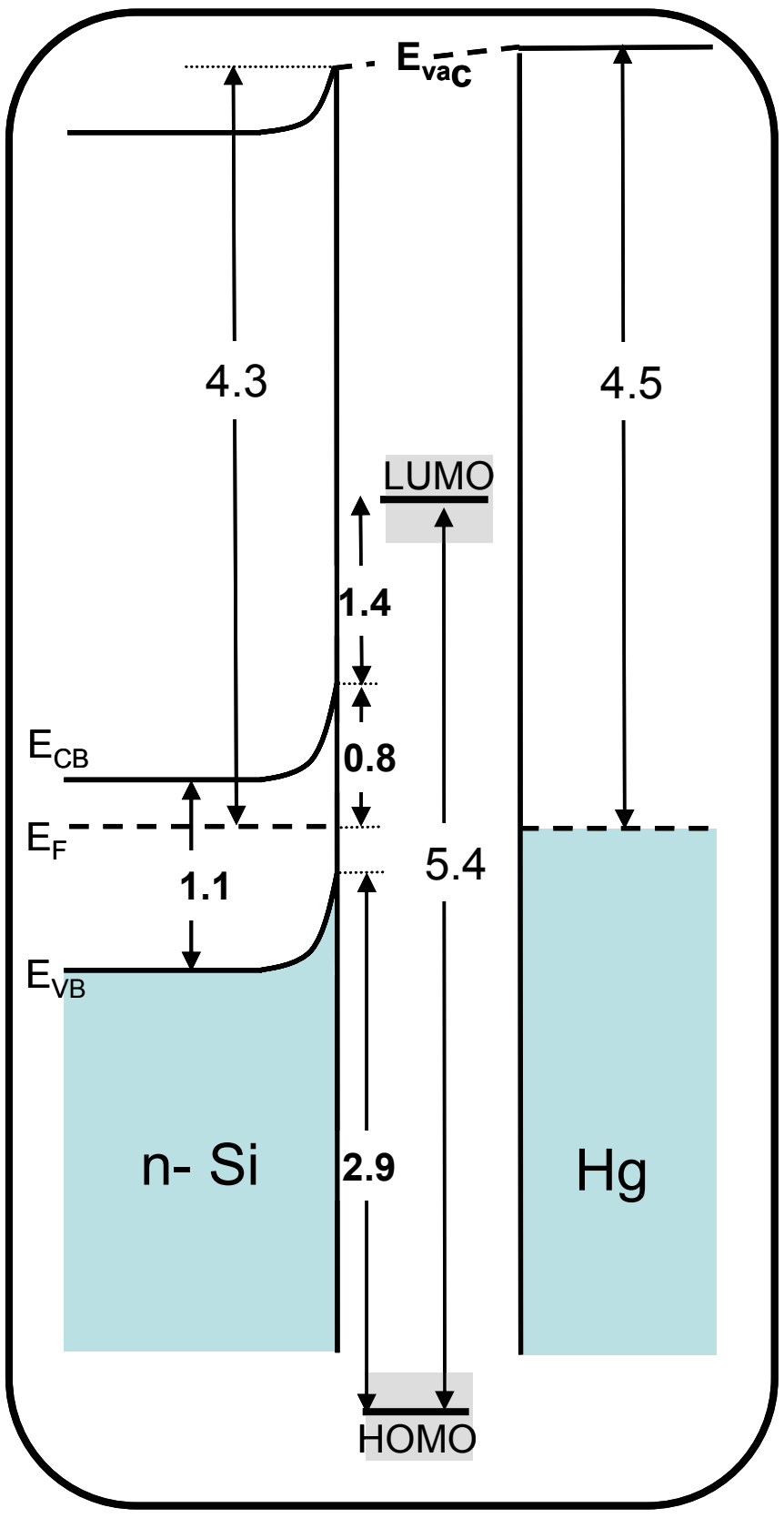

**Figure 3: Proposed band diagram for the n-Si-$C_{14}H_{29}$//Hg system from UPS, IPES, and J-V data. ($E_{CB}^{bulk}$ – $E_F$) = 0.25 eV. Experimental uncertainties in HOMO and LUMO values are indicated. The work functions are 4.3 eV(Si) and 4.5 eV(Hg). From UPS and XPS we find ~ 0.4 eV band bending before making Hg contact. The ~0.8 eV Schottky barrier implies that band bending increases to ~0.5 eV. As the LUMO was found to be ~2.2±0.3 eV above $E_F$, it is 1.4±0.3 eV above the Si $E_{CB}$.**

**Acknowledgements.**

We thank L. Kronik (WIS) and Y. Selzer & A. Nitzan (Tel Aviv Univ.) for helpful discussions and M. Gal (UNSW) for critical reading of the manuscript, the Clore Foundation for a fellowship to AS, the Israel Science Foundation (Jerusalem), Minerva Foundation (Munich) and the Philip M. Klutznick Research Fund for support to DC, the National Science Foundation (DMR-0408589) and the New Jersey Center for Organic Optoelectronics for support to AK, and the Australian Research Council for support to TB.